\documentclass{mn2e}
\usepackage{graphicx}
\title[A stellar merger]{A Carbon-rich Mira variable
in a globular cluster: A stellar merger
\thanks {Based on observations made with the Southern African Large Telescope SALT.}}
\author[Feast et al.]{ Michael W. Feast$^{1,2}$, John W. Menzies$^{2}$
and Patricia A. Whitelock$^{2,1}$\\
$^{1}$Astronomy, Cosmology and Gravity Centre, Astronomy Department, University of Cape Town,
7701, Rondebosch, South Africa \\
$^{2}$South African Astronomical Observatory, P.O. Box 9, 7935, Observatory, South Africa.}
\begin{document}
\maketitle
\begin{abstract}
The membership of Matsunaga's variable 1,  a carbon-rich, mass-losing, Mira variable, 
in the globular cluster Lynga 7 is discussed on the basis of radial velocities.
We conclude that it is a member, the first known C-Mira in a globular cluster.
Since such a variable is expected to have an age of $\sim 1-2$ Gyr and an
initial mass of $\sim 1.5$ solar masses, we conclude that this star must be the product of
a stellar merger.  
\end{abstract}
\begin{keywords}
stars:AGB and post-AGB - globular clusters;individual; Lynga 7 - stars; variable -
stars; carbon.
\end{keywords}
\section{Introduction}
In the course of an extensive infrared survey of Galactic globular clusters
for variable stars, Matsunaga (2006) discovered a long period Mira variable
near the centre of the globular cluster Lynga 7. The variable has a  period of 551 days,
a large infrared amplitude ($\Delta K = 1.22$ mag) and very red colours
(mean $(J-K) = 4.1$ mag) indicative of  mass-loss and a circumstellar dust shell.
Sloan et al. (2010)
obtained SPITZER mid-infrared spectra of the variable showing it to be a 
carbon star with strong  circumstellar dust emission from SiC and MgS particles as well as absorption from gaseous acetylene.  They estimate a mass loss of $2.5\times 10^{-7}$ 
solar masses per year. 
A few carbon rich objects are known in globular clusters, but these
have generally been interpreted as cluster examples of the CH stars in
the Galactic halo field. These latter stars are believed to be mass exchange binaries. However
no carbon Mira has previously been detected in a globular cluster and Matsunaga's
variable, V1, raises a number of problems related both to stellar evolution
and to the nature of Mira variables.

In the present paper we derive the radial velocity of V1 and discuss its cluster membership.
We then discuss the nature of V1 and its relation to the cluster population.

\section{Observations}
A medium dispersion (resolution 2.4A) spectrum of V1 Lynga 7 was obtained on
2012 May 11 (HJD 2456058.6) with the RSS spectrograph on the SALT telescope at SAAO, Sutherland. 
The spectral region covered was 6000 to 7100A. Spectra  of the
planetary nebula Hen 2-146 and the carbon Mira V650 CrA (P = 332 days)
were also obtained with the same equipment and settings.
 
Fig. 1 shows the spectrum  of V1 and that of V650 CrA. 
The spectra match each other well.
In both cases the spectrum is dominated by molecules of carbon compounds 
(particularly CN) and is typical of a carbon-rich Mira (see for instance
the atlas of Barnbaum, Stone \& Keenan (1996)).
In addition V1 shows strong $H\alpha$ emission. This is characteristic of C-Miras
at certain phases.
\begin{figure}
\includegraphics[width=8.5cm]{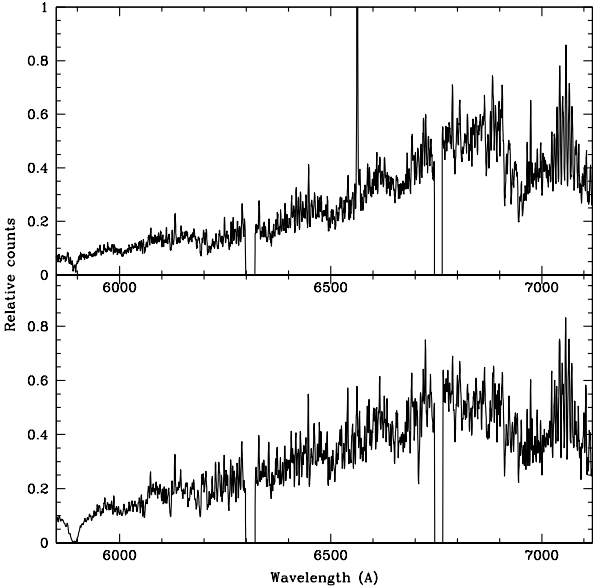}
\caption{Comparison of the Lynga7-V1 spectrum (upper panel) with that of the reference carbon star, V650 CrA (lower panel). The H$\alpha$ line in Lynga7-V1 has a peak count of 7880 e$^-$ (offscale at relative count of 1.64), while the peak count at ~7020$\AA$ in V650 CrA is 1.3x10$^5$ e$^-$. The counts go to zero in the gaps between the CCD chips.}
 \end{figure}

Using the comparison arc we obtain for
the planetary nebula Hen 2-146  a radial velocity
\footnote {Throughout we use heliocentric radial velocities.}
of $56\pm 4 \rm km\,s^{-1}$  
in good agreement with
the catalogue value of $61\pm 4 \rm km\,s^{-1}$ (Durand at al. 1998). 
Measured against the comparison arc the $H\alpha$ emission in V1 gives
$17 \rm km\,s^{-1}$ with an estimated uncertainty of $\sim 2 \rm km\,s^{-1}$.
The night sky lines in this spectrum yield $6 \pm 2 \rm km\,s^{-1}$. Thus we adopt
the $11 \rm km\,s^{-1}$ as our best estimate of the velocity from the $H\alpha$ emission.
This emission originates in 
shock waves in rising material in the stellar atmosphere. It therefore has, predominately,
a negative velocity with respect to the true radial velocity of the star. The best measure
(if available) of a C-Mira velocity is obtained from CO lines in the 
millimetre region, which arise
well above the photosphere
(see, e.g., Barnbaum 1992 and references therein). 
Data discussed in Menzies et al. (2006) show that on 
average Vel(CO(mm)) = Vel(Emission) + 16$\rm km\,s^{-1}$. Thus our best estimate of the
stellar velocity is $27 \rm km\,s^{-1}$. The uncertainty in this figure is not
easily determined, but is likely to be $5 \rm km\,s^{-1}$ or more.

A radial velocity was obtained for V1 by cross-correlating its spectrum with that of V650 CrA,
after masking out the strong $H\alpha$ emission line. 
A radial velocity of $3.5\rm km\,s^{-1}$ was adopted for V650 CrA (Walker 1979).
Several measures were made using
all or parts of the spectral range and we adopt a value of $40\pm10 \rm km\,s^{-1}$
as the best mean estimate. 
In Menzies et al (2006) an average  correction of 
$-4 \rm km\,s^{-1}$ of C-Mira absorption velocities to the CO (mm) standard system was
adopted, thus giving a estimate of $36 \pm 10\rm km\,s^{-1}$ for the stellar velocity.
The absorption line velocity of a C-Mira  varies in
a cycle by $\sim 13 \rm km\,s^{-1}$ (Sanford 1944). This will affect both 
V1 and V650 CrA and the cross correlation velocity is thus likely to be 
somewhat less certain than
that derived from $H\alpha$ emission. 

\section{Cluster membership of V1}
The position of V1 given by Matsunaga (2006) $(16^{h}11^{m}02^{s}.0 -55^{\circ}19^{'} 14{''} (J2000))$
together with the cluster centre given by Goldsbury et al. (2010) leads to a distance of the variable from the centre  of
$\Delta \rm (RA) =-13.5$ and $\Delta \rm (Dec) = -10$ arcsecs well within the core of the cluster (core radius 54 arcsecs, half-light radius 72 arcsecs (Harris 1996)).

Radial velocities of 9 cluster stars were measured by Saviane et al (2012).
They find a mean heliocentric velocity of $22\pm 3 \rm km\,s^{-1}$. Earlier
Tavarez \& Friel (1995) obtained $6 \pm 15 \rm km\,s^{-1}$ from 4 stars.
The high standard error in this case being due to an uncertainty in
night to night zero points. 
In section 2 we obtained a radial velocity for V1 from H$\alpha$ emission  
$27\pm \sim 5 \rm km\,s^{-1}$ with a 
somewhat less certain value of $36 \pm 10\rm km\,s^{-1}$ from
a match to V650 CrA. Given the uncertainties we consider
the agreement with the cluster velocity to be satisfactory.  

In addition to this, the velocity of V1 differs from that expected for
a field C-rich Mira of its period. Long period C-Miras in
the solar vicinity are disc objects showing evidence of differential
Galactic rotation. The velocity dispersions are moderate, 
$22\pm 4\rm km\,s^{-1}$ in the Galactic radial direction and
$18 \pm 3 \rm km\,s^{-1}$ in the direction of Galactic rotation,
at the relevant period.
(Feast et al. 2006). These dispersion include observational scatter
and may be slight overestimates. Assuming a flat rotation curve
($\Theta = 220 \rm km\,s^{-1}$ and $R_{0} = 8.0$kpc and with the local
solar motion from Feast \& Whitelock (1997), the predicted radial
velocity for V1 at the distance  predicted by a bolometric period-luminosity relation 
(see section 4) is $-81 \rm km\,s^{-1}$,
quite different from the value observed.
Also for a member of a disc population
with a relatively small velocity dispersion V1 is quite far from the 
Galactic plane at its predicted distance (445 pc).

\section{Discussion}
Saviane et al.(2012) determined a metallicity for Lynga 7 of [Fe/H] = $-0.57\pm 0.15$
on the scale of Carretta (2009) and
Sarajedini et al. (2007) show that an HST colour- magnitude diagram is
well fitted by that of 47 Tuc ([Fe/H] =$-0.76$). 
Lynga 7 has been claimed to be among the oldest globulars (Mar\'{i}n-Franch et al. 2009).
Sarajedini et al. estimate $(m-M)_{o} = 14.55$ and $E(B-V) = 0.78$
for Lynga 7. This is based 
on a fit to the 47 Tuc c-m diagram and
 adopting 13.40 and 0.055 for the latter.
There have been some differences in the distance
estimates for 47 Tuc (see Gratton et al. 2003)  and 
the uncertainty
in the modulus of Lynga 7 is probably not less than $\sim 0.1$mag.

For V1, Matsunaga obtained mean magnitudes of $J= 11.35$, $H=9.08$ and
$K_{s} = 7.25$ ($J-K_{s} = 4.10$) on the IRSF system which is close 
to the 2MASS system (Kato et al. 2007). Converted to the SAAO system using the transformations
given by Carpenter (2001 and web update) and
applying a reddening equivalent to $E(B-V) = 0.78$ on the Cardelli system (Cardelli
et al. 1989)
we obtain $m_{bol} = 9.60$ using the $BC_{K} -(J-K)$ relation
of Whitelock et al. (2006). A main uncertainty is probably the 
conversion from IRSF (assumed same as 2MASS) to SAAO.  
Thus with the Sarajedini distance modulus
the bolometric absolute magnitude of V1 is $-5.0$. The value predicted from the 
the period-luminosity $(M_{bol})$
relation of Whitelock et al.(2009) is $-5.2$. This relation is based on
LMC C-Miras with bolometric corrections calculated in the same way as for V1
and adopting a modulus of 18.5 for the LMC. 
Taking into account the various uncertainties including
the intrinsic scatter in the period-luminosity relation, the observed and predicted 
values are in satisfactory agreement.  
Note that V1 is 0.9 mag fainter at $K$
than predicted by a period-luminosity relation in $K$ 
(Whitelock et al. 2008), confirming that the star is
surrounded by an obscuring dust shell. 


Galactic kinematics (Feast et al. 2006) and, particularly, the presence of mass-losing
carbon Miras with periods $\sim 500$ days 
in intermediate age (1-2 Gyr) clusters in the Magellanic Clouds
(e.g. Nishida et al. 2000) strongly suggests that
the initial mass of a C-Mira such as V1 was about 1.5 solar masses.
V1 also matches properties of these cluster variables well in absolute
magnitude and colours.
On the other hand, Padova isochrones (Marigo et al. 2008, Giradi et al. 2010 and website) for 10-13Gyr globular clusters indicated that the initial mass
of their evolved stars is about 0.8 solar masses. 

Since the discovery of a carbon-rich star (a ``CH star") in $\omega$
Cen by Harding (1962) several such stars have been discovered. 
These are taken to be examples of mass exchange binaries of which
many are known in the general halo field. 
In such cases the amount of mass exchanged is likely to have
been relatively limited and sufficient only to pollute the stellar
atmosphere. Sloan et al. (2010) suggest that V1 could be a CH type
object. 
However, with a predicted initial mass of $\sim 1.5$ solar masses this
is not possible. The formation of a star of this mass
in Lynga 7 requires the merger of two stars near (or beyond)
the main sequence turn of point of the cluster. 

Sharina et al. (2012) have found a carbon star member in the 
globular cluster NGC 6426. It has quite strong molecular bands but is 
relatively blue in $JHK$. Nothing is known about its variability. 
The authors discuss whether it is  a mass exchange binary or a merger
and are not able to come to a certain conclusion (since the mass is
not known). A somewhat similar carbon rich star was found in the globular cluster
NGC6405 (M15) by Cot\'{e} et al. (1997), which they consider to be a typical CH star,
i.e. a binary with a white dwarf secondary which earlier donated mass to the observed star.

The nature of blue stragglers in the c-m of globular clusters has been
much discussed since they were first identified by Sandage (1953). It now 
seems that there are two types of blue stragglers, mass exchange binaries
and stellar mergers (see Ferraro et al. 2009 and
references therein). An initial discussion of
the evolution  of those blue stragglers which are merged stars
has been given by Sills et al. (2009). Further
theoretical work might allow one to determine whether the properties
of V1 require that the merger components be near the main sequence
or whether mergers of giant branch stars (whose internal chemical
composition has been changed by evolution, e.g. helium 
enrichment) would also result in a typical long period C-Mira.
It would be interesting to know whether there are blue stragglers in
Lynga 7. However, radial velocities or proper motions are required
for this (Milone et al. 2012). In any case objects like V1 must be very rare
since their life time as mass-losing AGB variables is quite short . 

 \section{Conclusions}
The radial velocity of Matsunaga's variable V1 is consistent with membership of the globular cluster Lynga 7.
Its absolute bolometric magnitude and colours are typical for a long period 
mass-losing carbon-rich
Mira variable. Since such 
variables have initial masses  of $\sim 1.5$ solar masses and ages of $\sim 1.5-2.0$Gyr,
we propose that V1 is the evolutionary product of a stellar merger in the
cluster $\sim 1.5$Gyr ago. 

\section*{Acknowledgements}
Each of the authors gratefully acknowledges the receipt  of research grants from
the National Research Foundation of South Africa (NRF).
The observations reported in this paper were obtained with the Southern
African Large Telescope, SALT.


\begin{thebibliography}{}
\bibitem[]{} Barnbaum C., 1992, ApJ, 385, 694
\bibitem[]{} Barnbaum C., Stone R.P.S., Keenan P.C., 1996, ApJSup, 105, 419
\bibitem[]{} Cardelli J.A., Clayton G.C., Mathis J.S., 1989, ApJ, 345, 245
\bibitem[]{} Carpenter J.M., 2001, AJ, 121, 2851
\bibitem[]{} Carretta E., Bragaglia A., Gratton R., D'Orazi V., Lucatello S., 2009, A\&A, 508,695
\bibitem[]{} Cot\'{e} P. et al. 1997, ApJ, 476, L15
\bibitem[]{} Durand S., Acker A., Zijstra A., 1998 A\&ASup, 136, 145
\bibitem[]{} Feast M.W., Whitelock P.A., 1997, MNRAS, 291, 683
\bibitem[]{} Feast M.W., Whitelock P.A., Menzies J.W., 2006, MNRAS, 369, 791
\bibitem[]{} Ferraro F.R. et al., 2009, Nature, 462, 1028
\bibitem[]{} Girardi L. et al., 2010, ApJ, 724, 1030
\bibitem[]{} Goldsbury R. et al., 2010, AJ, 140, 1830
\bibitem[]{} Gratton R.G. et al., 2003, A\&A, 408, 529
\bibitem[]{} Harding G.A., 1962, Observatory, 82, 205
\bibitem[]{} Harris W.E., 1996, AJ, 112, 1487
\bibitem[]{} Kato D. et al. 2007, PASJ, 59, 615
\bibitem[]{} Mar\'{i}n-Franch A. et al. 2009, ApJ, 694, 1498
\bibitem[]{} Marigo P. et al., 2008, A\&A 482,883
\bibitem[]{} Matsunaga N, 2006, PhD Thesis, University of Tokyo
\bibitem[]{} Menzies J.W., Feast M.W., Whitelock P.A., 2006, MNRAS, 369, 783
\bibitem[]{} Milone A.P et al., 2012, A\&A, 540, A16
\bibitem[]{} Nishida S. et al., 2000, MNRAS, 313, 136
\bibitem[]{} Sandage A.R., 1953, AJ, 58, 61
\bibitem[]{} Sanford R.F., 1944, PASP, 56, 237
\bibitem[]{} Sarajedini A. et al., 2007, AJ, 133, 1658
\bibitem[]{} Saviane I. et al., 2012, A\&A, 540, 27
\bibitem[]{} Sharina M., Aringer B., Davoust E., Kniazev A.Y., Donzelli C.J., 2012, ArXiv:
1207.4357
\bibitem[]{} Sills A., Karakas A., Lattanzio J., 2009, ApJ, 692, 1411
\bibitem[]{} Sloan G.C. et al., 2010, ApJ, 719, 1274
\bibitem[]{} Tavarez M., Friel E.D., 1995, AJ, 110, 223
\bibitem[]{} Walker A.R., 1979, South African Astron. Obs. Cir. 1, 112
\bibitem[]{} Whitelock P.A., Feast M.W., Marang F., Groenewegen M.A.T., 2006, MNRAS, 369, 751
\bibitem[]{} Whitelock P.A., Feast M.W., van Leeuwen F., 2008, MNRAS, 386, 313
\bibitem[]{} Whitelock P.A. et al. 2009, MNRAS, 394, 795
\end{thebibliography}
\end{document}